\documentclass{aa}  
\usepackage{graphicx}
\usepackage{amsmath}
\usepackage{txfonts}
\usepackage[]{hyperref}

\begin{document} 

   \title{gammapy$\_$SyLC}
   \subtitle{A Package for Simulating and Fitting Variability in High-Energy Light Curves}

   \author{C. Galelli
          \inst{1}
          }

   \institute{LUX, Observatoire de Paris, Université PSL, Sorbonne Université,
                CNRS, 92190 Meudon, France\\
              \email{galelli.cl@gmail.com}
             }
 
  \abstract
    {}
   {Characterizing the temporal variability of astrophysical sources is key to understanding the underlying physical processes driving their emissions. This work introduces a \texttt{gammapy$\_$SyLC}, a Python package that offers tools to simulate and fit time-domain data, with a focus on Active Galactic Nuclei (AGN) variability. The package was developed taking into account possible interactions with \texttt{gammapy} but does not directly depend on it.}
   {\texttt{gammapy$\_$SyLC} incorporates optimized implementations of the Timmer \& Koenig and Emmanoulopoulos algorithms for light curve simulation, capable of generating synthetic lightcurves from specified PSDs and amplitude distribution models. It also provides functionalities for PSD fitting, histogram-based PDF interpolation, and Monte Carlo-based parameter estimation, making it a full-stack tool for investigating variable phenomena and specifically the long-term behavior of AGNs. }
   {To showcase its capabilities, the package was applied to gamma-ray light curves from the Fermi Large Area Telescope repository, reconstructing PSDs and PDFs and constraining variability models for observed sources.}
   {}


   \maketitle
%

\section{Introduction}

The temporal behavior of astrophysical sources provides crucial insights into the underlying physical processes that govern their emission mechanisms. Time-domain studies are fundamental in advancing the understanding of phenomena such as accretion, relativistic jet formation, and particle acceleration in high-energy astrophysical systems. For active galactic nuclei (AGNs), variability across a wide variety of timescales, from hours to decades, offers a unique lens through which to observe the interplay between the central supermassive black hole, the accretion disk, jets, blobs of plasma, and surrounding emission regions. Notably, the stochastic nature of AGN variability reflects the complex physical processes that shape the emitted radiation, making quantitative time-series analysis essential for deep-scale interpretation.

In the gamma-ray regime, the advent of sensitive instruments like the Fermi Large Area Telescope (Fermi-LAT) has significantly enriched the field of time-domain astronomy. By providing long-term, high-cadence monitoring of AGNs, Fermi-LAT has enabled detailed studies of variability, offering important insights on many phenomena and observables, such as flaring events, periodic behavior, and power spectral density (PSD) slopes. However, the effective interpretation of such data often requires advanced tools capable of simulating, fitting, and analyzing the statistical properties of the observed light curves.

This paper introduces \texttt{gammapy$\_$SyLC}\footnote{\url{https://github.com/cgalelli/gammapy_SyLC}}, a Python package designed to address these challenges by complementing existing analysis tools such as \texttt{gammapy} (\citet{donath2023}). While not dependent on \texttt{gammapy}, \texttt{gammapy$\_$SyLC} integrates seamlessly with its functionalities, providing a specialized set of algorithms for time-domain studies. Central to the package are implementations of the Timmer \& Koenig (\citet{timmer1995})and Emmanoulopoulos (\citet{emmanoulopoulos2013}) algorithms, optimized for simulating light curves with specified PSDs and flux amplitude distributions. Additionally, the package includes tools for reconstructing empirical probability density functions (PDFs), generating PSD and histogram envelopes, and performing Monte Carlo-based parameter fitting.

The goal of \texttt{gammapy$\_$SyLC} is to bridge the gap between theoretical models and observed variability, enabling researchers to simulate realistic light curves, fit statistical models, and extract meaningful constraints on source properties. To illustrate its capabilities, the package was applied to gamma-ray light curves from the Fermi-LAT light curve repository (LCR)\footnote{Openly available at \url{https://fermi.gsfc.nasa.gov/ssc/data/access/lat/LightCurveRepository/}}, demonstrating its effectiveness in reconstructing PSDs and PDFs and constraining the long-term variability models of AGNs.

The remainder of this paper is organized as follows. Section 2 describes the simulation algorithms implemented in \texttt{gammapy$\_$SyLC}, focusing on their ability to generate realistic light curves. Section 3 discusses the package's fitting functionalities, emphasizing its Monte Carlo approach to parameter estimation. Finally, Section 4 presents results obtained by applying these methods to real data, showcasing the package's practical utility in gamma-ray time-domain studies.


\section{Simulation Algorithms}

\begin{figure*}
   \resizebox{\hsize}{!}{\includegraphics[clip]{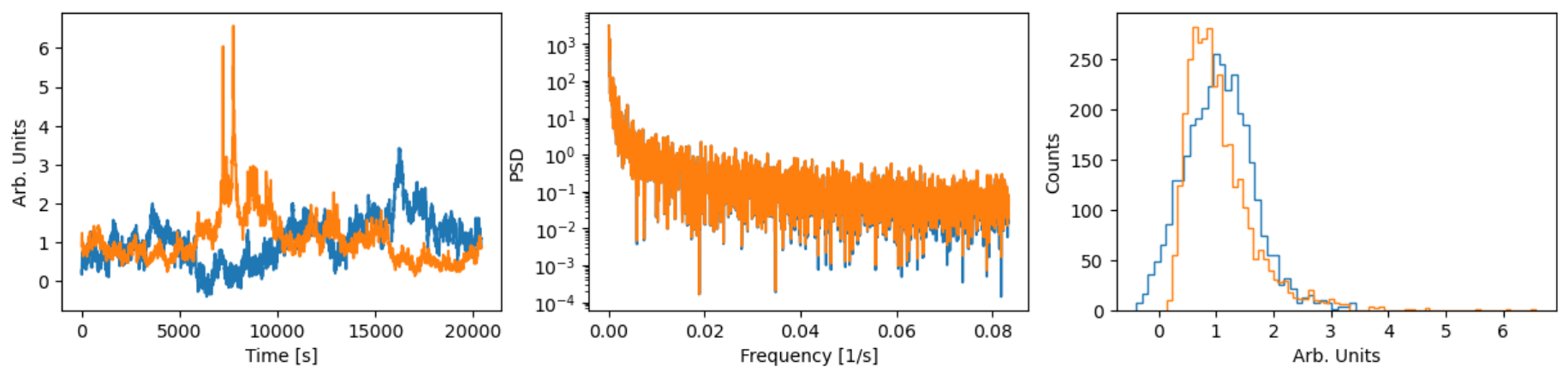}}
    \caption{Left: Simulated time series using the TK and EMM algorithms. Center: Periodograms of the light
curves. Right: Amplitude distribution of the light curves. TK in blue, EMM in orange.}
         \label{tkvsemm}
\end{figure*}

Simulating light curves with desired variability properties is a fundamental tool in time-domain astrophysics. It enables the study of variability patterns, the testing of observational models, and the derivation of physical interpretations from temporal data. \texttt{gammapy\_SyLC} includes two algorithms for generating synthetic light curves: the Timmer-König (TK) algorithm and the Emmanoulopoulos (EMM) algorithm.

\subsection{The Timmer-König Algorithm}
The Timmer-König algorithm generates Gaussian-distributed light curves that match a specified power spectral density (PSD) model. The PSD describes the variability as a function of frequency as a description of the temporal behavior of the light curve in the Fourier space. 

The algorithm operates by randomizing the phases and amplitudes of Fourier components derived from the target PSD. The inverse Fourier transform of these components yields a time series with the desired frequency-domain properties. However, this method inherently produces normally distributed data, which may not accurately represent astrophysical sources, which often exhibit non-Gaussian flux distributions, especially in the high-energy domain in which statistics are scarce.

\subsection{The Emmanoulopoulos Algorithm}
The Emmanoulopoulos algorithm extends the capabilities of the TK method by allowing the generation of light curves with a specified amplitude distribution (PDF), in addition to the power spectral model. This feature is critical for simulating astrophysical sources with skewed or heavy-tailed flux distributions, such as AGNs.

The algorithm follows a four-step process:
\begin{enumerate}
    \item A Gaussian-distributed light curve is created using the TK method, matching the target PSD, and its Fourier transform is computed.
    \item From the PDF model, a white-noise (WN) series of pseudo-random numbers with the same length as the TK curve is produced, and its Fourier transform is computed.
    \item For each frequency, the amplitudes of the WN transform are replaced with the amplitudes of the TK transform while keeping the phases unaltered. A new time series (ADJ) is obtained by inverse Fourier transform.
    \item A new time series is created from the values of WN ordered based on the ranking of ADJ.
\end{enumerate}
These steps are then repeated by replacing WN with the final resulting series. In the \texttt{gammapy\_SyLC} implementation, the iterative process halts when changes between successive iterations fall below a threshold, or reach a maximum of iterations, ensuring efficiency without compromising accuracy.

\subsection{Implementation notes}

In \texttt{gammapy\_SyLC}, the implementations of both algorithms are adapted by using strategies that improve the result or make it more malleable to the user’s request. These include:
\begin{itemize}
    \item \textbf{Oversampling}: The algorithm generates an extended time series (controlled by the parameter \texttt{nchunks}, defaulting to 10). The final light curve is extracted from the center of this extended series to minimize edge effects and mitigate spectral leakage.
    \item \textbf{Flux Rescaling}: Users can specify the mean (\texttt{mean}) and standard deviation (\texttt{std}) of the output light curve. By default, the mean is set to 0 and the standard deviation to 1.
    \item \textbf{Noise Component}: Gaussian or Poissonian noise can be introduced, governed by the parameters \texttt{noise} and \texttt{noise\_type}, to simulate measurement uncertainties or intrinsic noise properties.
\end{itemize}

\subsection{Comparison and Use Cases}
The choice between the TK and EMM algorithms depends on the scientific goals. For cases where the amplitude distribution is unimportant, the TK algorithm offers a straightforward and computationally efficient solution. For studies requiring realistic flux distributions, the EMM algorithm is indispensable. A comparison of the results is shown in figure \ref{tkvsemm}, where it is evident that two curves, produced by the two algorithms with the same seed, differ in the flux distribution but have identical periodogram. 

Both algorithms are implemented with multiprocessing support, allowing users to efficiently generate ensembles of light curves for Monte Carlo simulations. The outputs can be directly used for PSD fitting, statistical analysis, or as inputs for further modeling.


\begin{figure*}
   \centering
   \includegraphics[width=0.8\hsize]{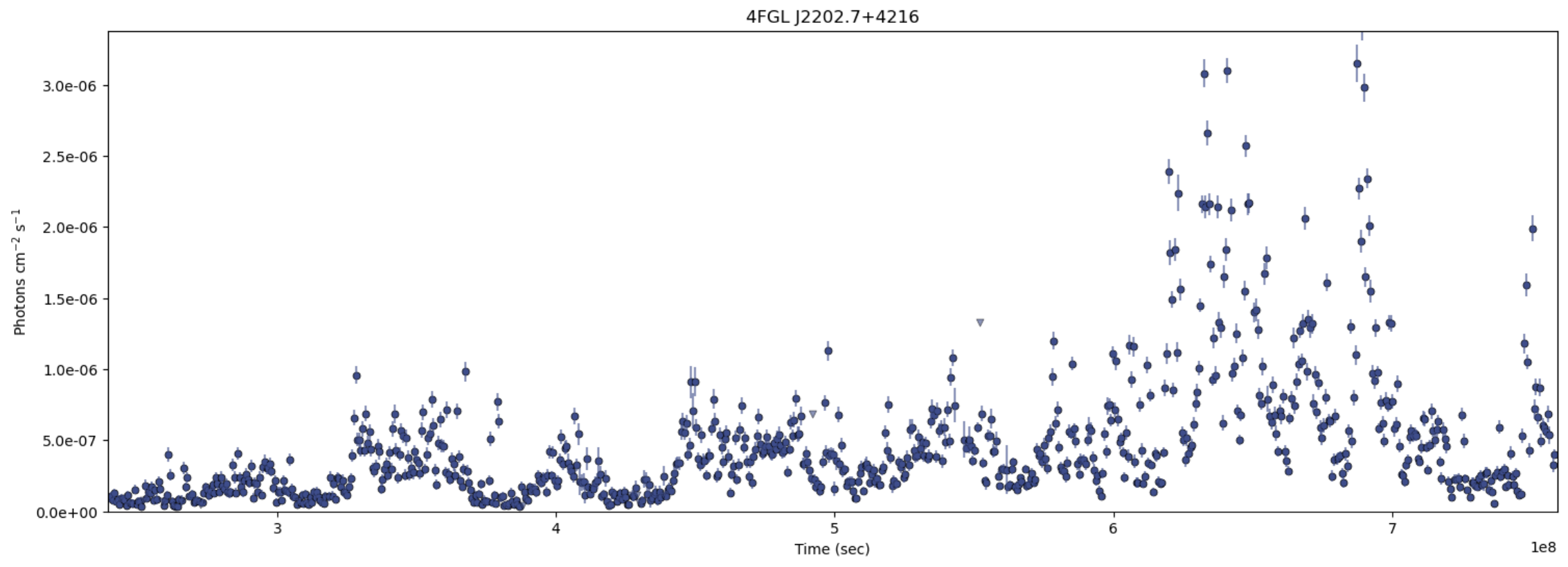}
      \caption{Weekly-cadenced lightcurve for BL Lac}
         \label{curve}
\end{figure*}

\section{Fitting Procedures in \texttt{gammapy\_SyLC}}

Fitting the observed properties of a light curve to theoretical models is a cornerstone of time-domain astrophysics. \texttt{gammapy\_SyLC} provides robust and flexible implementations for fitting both the power spectral density (PSD) and probability density function (PDF) of flux amplitudes. The integration of multiprocessing in both procedures allows users to efficiently perform large-scale simulations, making it practical for high-cadence or long-duration light curves.

\subsection{Power Spectral Density (PSD) Fitting}

The PSD fitting functionality in \texttt{gammapy\_SyLC} is designed to estimate the parameters of a theoretical PSD model by comparing it to an observed periodogram. This process involves generating synthetic light curves that follow the target PSD, constructing statistical envelopes of their periodograms, and optimizing the model parameters to match the observed data.

The PSD fitting procedure is based on the following steps:
\begin{enumerate}
    \item \textbf{Simulated Periodograms:} Synthetic light curves are generated using the Timmer \& Koenig or Emmanoulopoulos algorithms, depending on whether the PDF of flux amplitudes needs to be considered. Each simulated light curve is transformed into the frequency domain to compute its periodogram.
    \item \textbf{Envelope Construction:} From multiple simulated periodograms, statistical envelopes are constructed, representing the expected mean PSD and its statistical variability across simulations.
    \item \textbf{Goodness-of-Fit Calculation:} The observed periodogram is compared to the simulated envelopes using a custom goodness-of-fit metric. \texttt{gammapy\_SyLC} employs a $\chi^2$-like statistic, weighted on the number of simulated instances featuring a better $\chi^2$ score than the observed data.
    \item \textbf{Parameter Optimization:} The parameters of the theoretical PSD model are adjusted using \texttt{scipy.optimize.minimize} \footnote{\url{https://docs.scipy.org/doc/scipy-1.15.0/reference/optimize.html}} (\citep{virtanen2020}) to minimize the goodness-of-fit statistic.
\end{enumerate}

Monte Carlo simulations are performed to estimate uncertainties in the fitted parameters. This involves repeating the fitting process with multiple realizations of the observed periodogram to quantify the variability in the parameter estimates and building a confidence interval. This technique is known as \textit{Neyman construction} (\citep{neyman1937}).

\subsection{Probability Density Function (PDF) Fitting}

The PDF fitting procedure is designed to characterize the flux amplitude distribution of a light curve, focusing on its statistical properties. This is particularly useful for systems where the amplitude distribution deviates significantly from Gaussian, such as AGNs with heavy-tailed or skewed flux distributions, which can highlight complex aspects of the emission mechanisms, complementary to the ones observable by looking at the periodogram.

\renewcommand{\arraystretch}{1.5}
\begin{table*}[h!]
\centering
\begin{tabular}{| l | c c c c |}
\hline
\textbf{Object Name} & \textbf{Spectral Index (PSD)} & \textbf{P(Lognormal/Normal)} & \textbf{P(Lognormal/Gamma)} & \textbf{P(Lognormal/Alpha-Stable)} \\
\hline \hline
BL Lac & -0.91$\pm$0.04  & 0.03 & 0.13 & 0.75 \\
\hline
\end{tabular}
\caption{Fitting results for BL Lac.}
\label{tab:results_bllac}
\end{table*}

The PDF fitting algorithm employs the following steps:
\begin{enumerate}
    \item \textbf{Simulated Light Curves:} Light curves are generated using the Emmanoulopoulos algorithm to ensure that the PSD and PDF properties match the target models.
    \item \textbf{Interpolated PDF:} A piecewise cubic Hermite interpolating polynomial (PCHIP) is constructed from the flux amplitudes of the simulated light curves. This interpolated PDF captures the empirical distribution of the flux values.
    \item \textbf{Likelihood Calculation:} The negative log-likelihood (NLL) is computed by evaluating the observed flux amplitudes against the interpolated PDF. Uncertainties in the flux are taken into account by extracting $N$ random samples with a mean equal to the flux measurement and standard deviation equal to the uncertainty and taking the geometrical average of the likelihoods of the random samples. The final form of the NLL is then:
    \begin{equation*}
    \text{NLL} = -\sum_{i=0}^n \log\left( \prod_{j=0} ^m \text{PDF}_{interp} \left( \mathcal{N}_j(x_i, xerr_i) \right) \right)^{\frac{1}{m}}
    \end{equation*}
    \item \textbf{Parameter Optimization:} The PDF parameters are optimized using the \texttt{scipy.optimize.minimize} function. The optimizer minimizes the NLL to find the best-fit parameters.
\end{enumerate}

To further evaluate the robustness of the fitted PDF, \texttt{gammapy\_SyLC} provides two statistical tests:
\begin{itemize}
    \item Normality-based likelihood ratio test (\texttt{test\_norm}): This function assesses the significance of the PDF fit by comparing the likelihood obtained from the real data to a distribution of likelihood values obtained from synthetic light curves generated under the assumption of a normally distributed flux. Using the TK method, synthetic light curves are simulated and fitted with the same PDF model as the real data, and the difference in likelihood values is computed across multiple trials. The fraction of synthetic realizations that yield a worse fit than the real data provides a measure of how well the assumed PDF explains the observations.
    \item Alternative PDF Model Comparison (\texttt{test\_models}): This function follows a similar methodology but instead tests the fitted PDF against light curves generated with an alternative amplitude distribution. Instead of assuming a Gaussian process, synthetic light curves are generated using the EMM algorithm with an alternative PDF model. By fitting the test PDF to these synthetic light curves and computing the difference in likelihood, it is possible to determine whether the assumed model is significantly better than the alternative.
\end{itemize}

These tests provide additional validation layers for the selected amplitude distribution model, ensuring the fitted PDF well represents the observed flux variations.

\section{Application to Fermi-LAT lightcurves}

The Fermi Large Area Telescope (Fermi-LAT) is a space-based gamma-ray telescope that has provided long-term, high-cadence monitoring of astrophysical sources since its launch in 2008. With its broad energy range (20 MeV to over 300 GeV) and all-sky survey mode, Fermi-LAT has been instrumental in studying the variability of AGNs and other high-energy astrophysical phenomena. The Fermi-LAT Light Curve Repository (LCR) offers preprocessed light curves for numerous gamma-ray sources, facilitating detailed time-domain studies. For the analysis, the interface with the LCR was performed by the \texttt{pyLCR} python package, by Daniel Koncevski\footnote{\url{https://github.com/dankocevski/pyLCR}}. An example of a light curve obtained from this interface, for the AGN 4FGL J2202.7+4216 (BL Lac), is shown in figure \ref{curve}.

\subsection{Procedure}

The light curves of the Fermi-LAT LCR were analyzed using the functionalities provided by \texttt{gammapy\_SyLC}. The analysis consisted of the following steps:

\begin{enumerate}
    \item \textbf{Target Selection:} The AGNs selected for this study represent a diverse and well-characterized sample of gamma-ray variable sources, allowing for a comprehensive investigation of variability properties. BL Lac \citep{Ackermann2011, Raiteri2013, Madejski1999} was chosen as a reference source due to its historical significance and extensive observational coverage, making it an ideal test case for power spectral density and probability density function fitting techniques. Markarian 421 \citep{Abdo2011, Acciari2020} and Markarian 501 \citep{Abdo2011_Mrk501, Ahnen2017, Albert2007} are two of the closest and brightest BL Lac objects. PG 1553+113 \citep{Ackermann2015, Aleksic2015} was included due to its quasi-periodic behavior observed in gamma rays. The inclusion of 3C 279 \citep{Hayashida2015, Ackermann2016}, 4C 01.02 \citep{Massaro2015, Nolan2012}, and CTA 102 \citep{Gasparyan2018, Ahnen2018, Shukla2018} ensures representation of flat-spectrum radio quasars (FSRQs), which exhibit different variability properties compared to BL Lac objects.
    
    \item \textbf{Data Acquisition:} Light curves were retrieved from the Fermi-LAT Light Curve Repository (LCR), containing flux measurements, associated uncertainties, and test statistics. Datasets were downloaded with weekly cadence excluding points with non-convergence status from the Fermi analysis.

    \item \textbf{Periodogram fitting:} The periodogram was estimated from the observed light curve. The theoretical PSD was modeled using a simple power-law function, and its index was fitted using the PSD fitting algorithm, also reconstructing the uncertainty on the parameter.

    \item \textbf{Flux Distribution Modeling:} The probability density function of the observed flux amplitudes was analyzed to assess the underlying statistical properties of variability. A lognormal distribution was initially assumed, and its validity was tested against a normal distribution. For selected sources, additional comparisons were performed against alternative models, namely the gamma distribution and the maximal skewed alpha-stable distribution \citep{mcculloch1986, Biteau2012}, to evaluate the presence of heavy-tailed or strongly asymmetric flux distributions.
\end{enumerate}

\renewcommand{\arraystretch}{1.5}
\begin{table*}[h!]
\centering
\begin{tabular}{| l | c c c |}
\hline
\textbf{Object Name} & \textbf{Spectral Index (PSD)} & \textbf{P(Lognorm/Norm)} & \textbf{P(Lognorm/Gamma)} \\
\hline \hline
Mkn 421 & -0.83$\pm$0.05  & 0.0014 & 0.08 \\ 
\hline
Mkn 501 & -0.75$\pm$0.04  & 0.45 & 0.78 \\
\hline
\end{tabular}
\caption{Fitting results for Markarian 421 and Markarian 501.}
\label{tab:results_mkn}
\end{table*}

\subsection{Discussion}

The analysis of gamma-ray light curves measured by Fermi-LAT from different AGNs using \texttt{gammapy\_SyLC} reveals diverse variability characteristics among the selected sources. While some AGNs exhibit clearer signatures in their PSD and PDF fits, others show weaker variability at weekly cadences, making detailed statistical modeling more challenging. Here, the results for each source are summarized.

\subsubsection{BL Lac}

\begin{figure}
   \centering
   \includegraphics[width=\hsize]{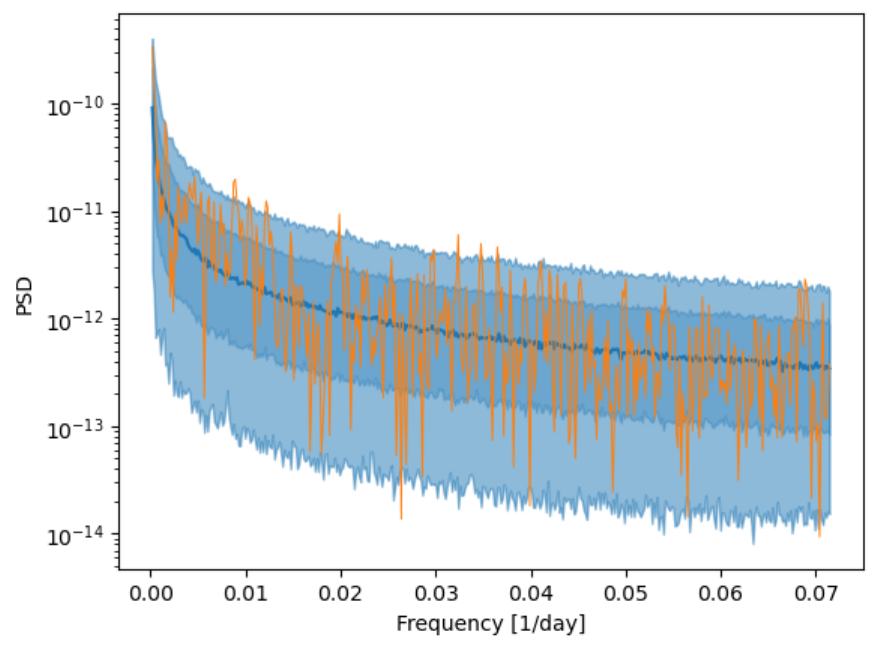}
   \caption{Periodogram of BL Lac overlaid with the best-fit PSD model and 68\% and 95\% quantile envelopes.}
   \label{fig:psd_bllac}
\end{figure}

BL Lac's extreme variability is reflected both in the power spectral density and the amplitude distribution. The PSD fitting returns a power-law index of -0.91 with an uncertainty of approximately 5\%, confirming the typical red-noise behavior observed in previous studies of BL Lac-type objects. The significant variability observed at all frequencies enhances the power-law shape of the PSD, effectively overpowering the white noise component, and leading to a clear reconstruction of the intrinsic temporal behavior of the source. This makes BL Lac an excellent example candidate for PSD modeling. The observed periodogram is plotted against the 1 and 2$\sigma$ statistical envelopes of the best-fit PSD model in figure \ref{fig:psd_bllac}. 

\begin{figure}
   \centering
   \includegraphics[width=\hsize]{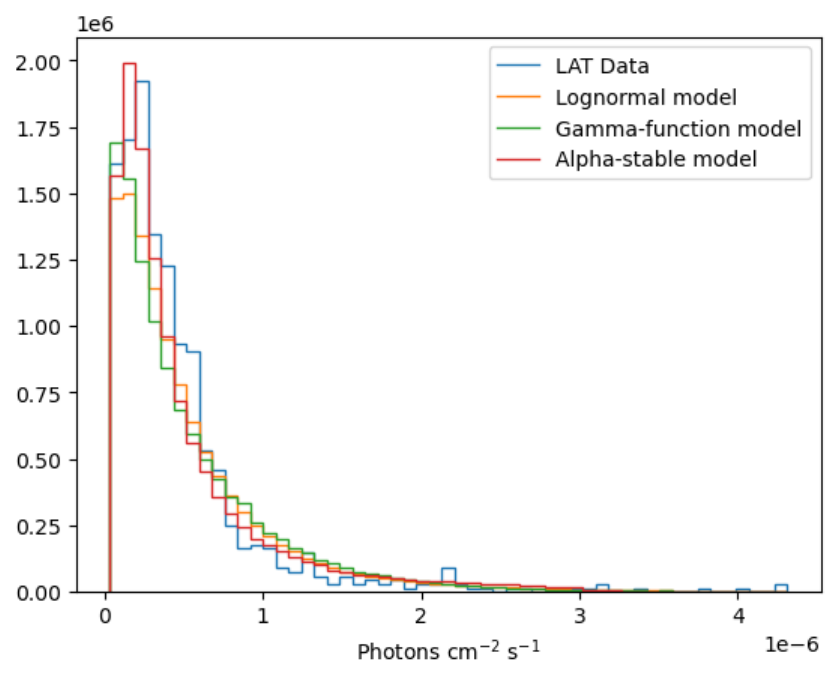}
   \caption{Flux distribution of BL Lac compared to the best-fit lognormal, gamma, and alpha-stable models.}
   \label{fig:pdf_bllac}
\end{figure}

However, the same extreme variability that facilitates the PSD fitting complicates the amplitude distribution analysis. The highly skewed distribution with a long tail makes it difficult to fit the shape with a simple PDF model. Even with nearly 1000 weekly flux points, the statistical power to distinguish between different models is limited. In particular, the lognormal distribution shows a moderate preference over the normal distribution at a significance level of 2$\sigma$; this non-decisive preference is more a factor of the non-optimality of the lognormal model rather than the normality of the distribution. The comparison with the gamma distribution and the maximal skewed alpha-stable distribution yield lower significance values, well below 2$\sigma$. In particular, the comparison with the alpha-stable distribution yields that the two models are much more similar to each other than to data, rendering the test completely inconclusive. This highlights that the complexity of the amplitude distribution, with its heavy tail and asymmetry, cannot be fully captured by the currently tested models. Figure~\ref{fig:pdf_bllac} illustrates the observed flux distribution against the three best-fit PDF models, emphasizing the challenges in accurately modeling the extreme tails.

Overall, the PSD analysis provides good information on the temporal variability of BL Lac, while the PDF fitting reveals the limitations of simple distribution models for highly variable sources. These findings underline the need for either more sophisticated models or larger datasets to improve the discrimination power among competing amplitude distribution hypotheses.

A summary of the results of the different analyses on BL Lac is shown in table \ref{tab:results_bllac}.

\subsubsection{Markarian 421 and Markarian 501}

Markarian 421 and Markarian 501 display different variability characteristics compared to BL Lac, which impacts the results of the PSD and PDF analyses. 

\begin{figure}
   \centering
   \includegraphics[width=\hsize]{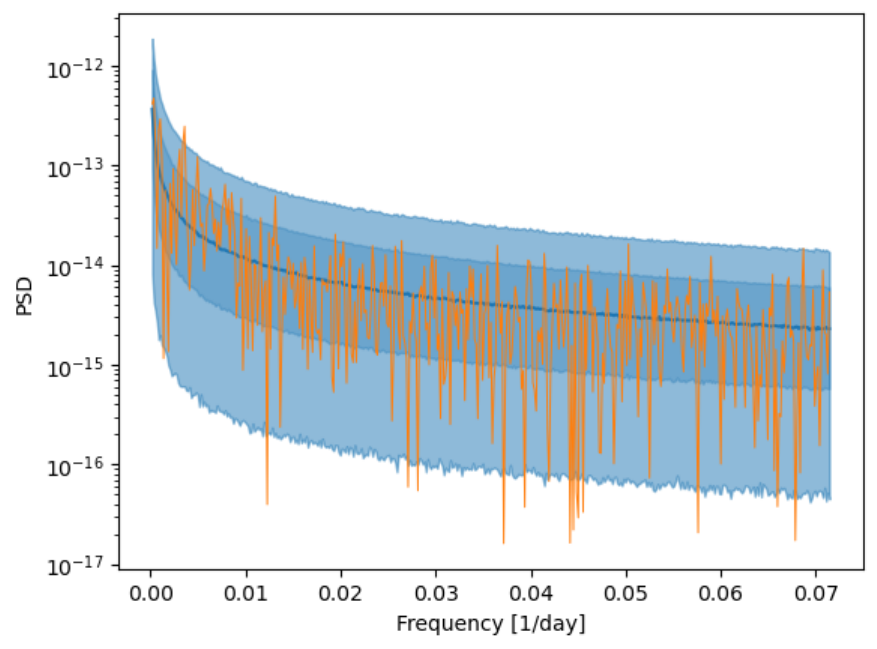}
   \caption{Periodogram of Markarian 421 overlaid with the best-fit PSD model and 68\% and 95\% quantile envelopes.}
   \label{fig:psd_mrk421}
\end{figure}

\begin{figure}
   \centering
   \includegraphics[width=\hsize]{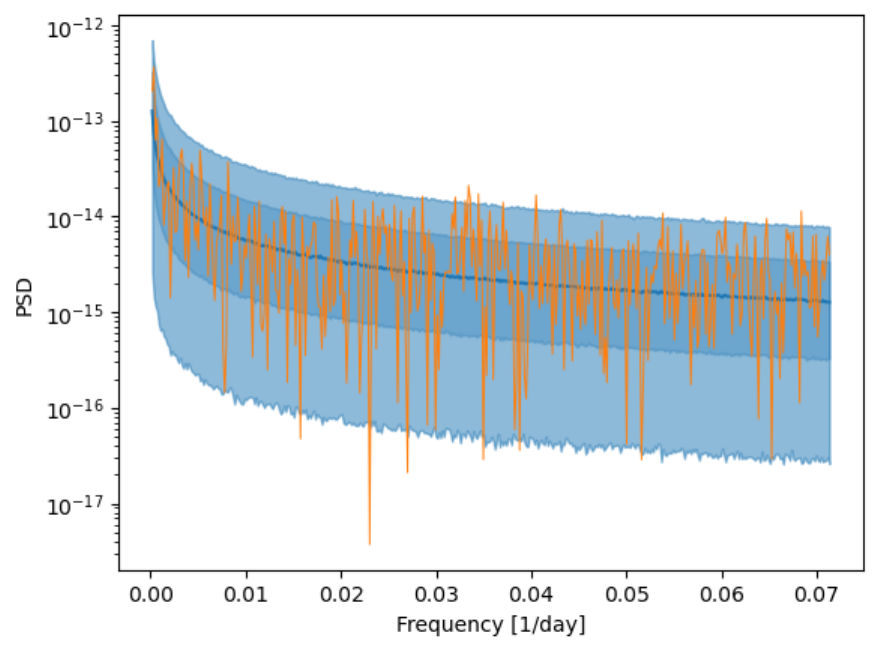}
   \caption{Periodogram of Markarian 501 overlaid with the best-fit PSD model and 68\% and 95\% quantile envelopes.}
   \label{fig:psd_mrk501}
\end{figure}

For the PSD fitting, both sources show a red noise component consistent with a power-law behavior, with best-fit spectral indices of -0.83 for Markarian 421 and -0.75 for Markarian 501. However, unlike BL Lac, a clear white noise component can be seen by the eye in the high-frequency region of the periodogram for both objects. This white noise component introduces a plateau in the periodogram, reducing the accuracy of the single power-law index fit and contributing to a slightly flatter reconstructed slope. The observed periodograms for both sources are shown in Figures~\ref{fig:psd_mrk421} and~\ref{fig:psd_mrk501}, where the best-fit power-law models and statistical envelopes are overlaid.

\begin{figure}
   \centering
   \includegraphics[width=\hsize]{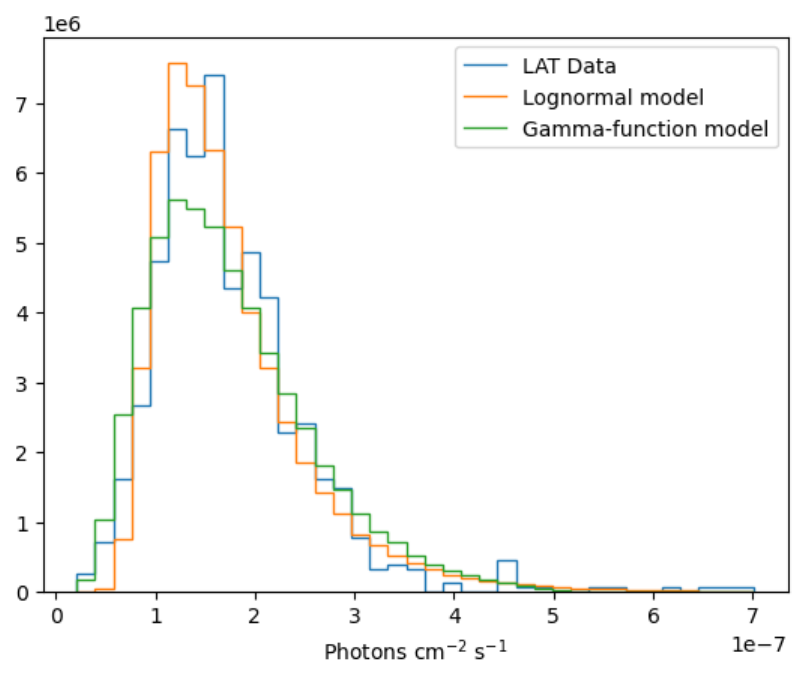}
   \caption{Flux distribution of Markarian 421 compared to the best-fit lognormal and gamma models.}
   \label{fig:pdf_mrk421}
\end{figure}

In terms of amplitude distribution, Markarian 421 and Markarian 501 exhibit distinct behaviors. By eye, Markarian 421 shows a moderately skewed flux distribution, clearly deviating from a simple Gaussian shape. In fact, the lognormal model provides a good fit to the data when compared with a Gaussian hypothesis with a significance level of 3$\sigma$, indicating a preference for the former. However, the comparison between the lognormal and gamma distributions remains inconclusive. This demonstrates that, even in cases of less extreme skewness, the discrimination power between similar-tailed distributions remains weak. The histogram of flux amplitudes for Markarian 421 and the best-fit PDFs for lognormal and gamma distributions are shown in Figure~\ref{fig:pdf_mrk421}. 

\begin{figure}
   \centering
   \includegraphics[width=\hsize]{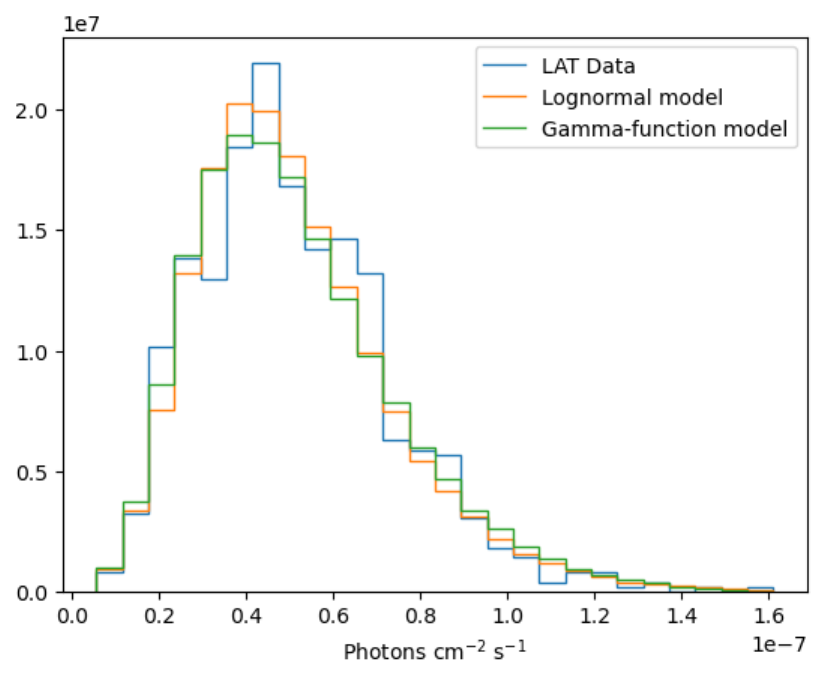}
   \caption{Flux distribution of Markarian 501 compared to the best-fit lognormal and gamma models.}
   \label{fig:pdf_mrk501}
\end{figure}

 Markarian 501 displays a more Gaussian-like flux distribution. This is consistent with its PSD, which exhibits a flatter slope compared to Markarian 421. The predominance of short-term variability, which is typically less extreme and more Gaussian-like, results in an amplitude distribution more closely resembling a normal distribution. Consequently, the separation between lognormal and normal models is significantly less robust. Similarly, the comparison between lognormal and gamma distributions is inconclusive, reflecting the challenges of distinguishing between these models for nearly symmetric amplitude distributions. Figure~\ref{fig:pdf_mrk501} illustrates the histogram and fitted PDFs for Markarian 501, showing the close overlap between the different models. 

The overall results for the two objects are shown in table \ref{tab:results_mkn}.

\subsubsection{PG 1553+113}

\begin{figure}
   \centering
   \includegraphics[width=\hsize]{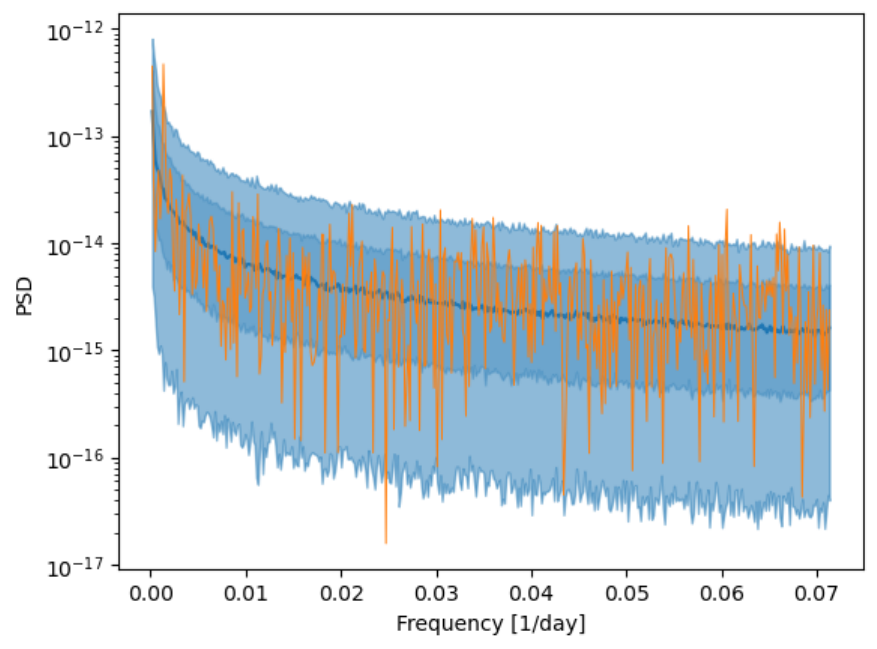}
   \caption{Periodogram of PG1553+113 overlaid with the best-fit PSD model and 68\% and 95\% quantile envelopes.}
   \label{fig:psd_pg1553}
\end{figure}

The PSD analysis of PG 1553+113 reveals a power-law behavior with a spectral index of $-0.78 \pm 0.03$.  Previous studies have suggested for this object a quasi-periodic oscillation (QPO) at a period of approximately 2.18 years \citep{Ackermann2015}. While the simple power-law reconstruction employed here is not explicitly designed to detect periodic signals, it effectively models the overall variability trend and recovers the spectral index. Interestingly, the periodogram value corresponding to the frequency closest to the proposed QPO (2.04 years in the sampled frequency) exceeds both the 2$\sigma$ and 3$\sigma$ confidence envelopes for the best-fit power-law model. This protrusion suggests an excess of power at this frequency, consistent with the hypothesized periodic behavior. Additionally, a flattening of the periodogram is observed at higher frequencies, indicating the dominance of white noise beyond a certain frequency in a similar fashion to the Markarians, and the slight reduction in the spectral index value somewhat masks the amplitude of the periodic peak, making the QPO feature less distinguishable. The fitted model and the statistical envelopes are shown in figure~\ref{fig:psd_pg1553}.

\begin{figure}
   \centering
   \includegraphics[width=\hsize]{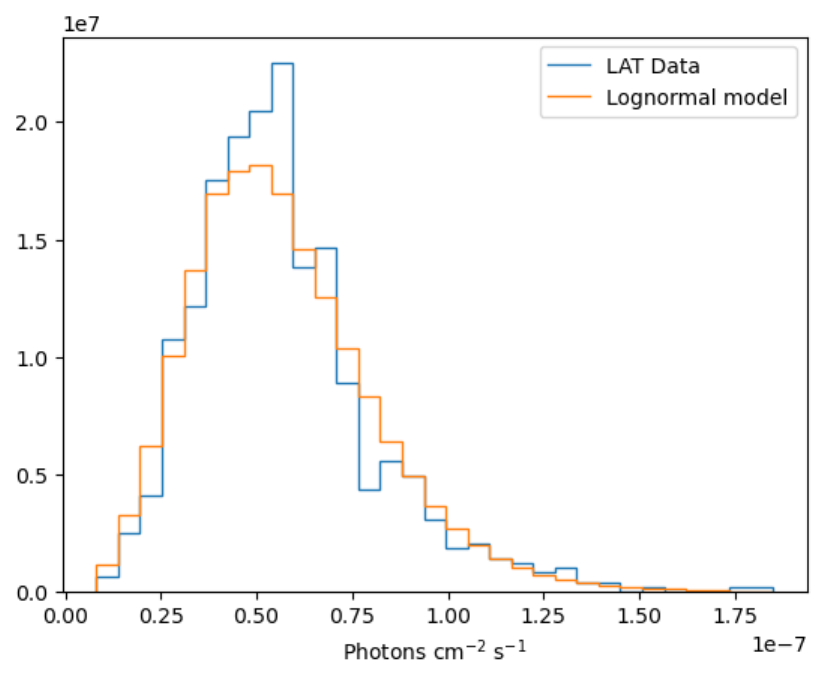}
   \caption{Flux distribution of PG1553 compared to the best-fit lognormal model.}
   \label{fig:pdf_pg1553}
\end{figure}

Regarding the amplitude distribution, PG 1553+113 exhibits a shape similar to that of Markarian 501. The histogram of flux values shows a very moderately skewed distribution, slightly deviating from normality but not enough to favor a lognormal model - the test shows below-1$\sigma$ preference. This lack of strong model separation can be attributed to the relatively mild skewness of the flux distribution, which complicates the discrimination between the tested models. The fitted PDFs are shown in figure \ref{fig:pdf_pg1553}.

\subsubsection{Flat Spectrum Radio Quasars: 3C 279, CTA 102, 4C 01.02}

\renewcommand{\arraystretch}{1.5}
\begin{table}[h!]
\centering
\begin{tabular}{| l | c c |}
\hline
\textbf{Object Name} & \textbf{Spectral Index (PSD)} & \textbf{P(Lognorm/Norm)} \\
\hline \hline
3C 279 & -0.84$\pm$0.03  & 0.11 \\ 
\hline
CTA 102 & -0.93$\pm$0.04  & 0.14 \\
\hline
4C 01.02 & -1.01$\pm$0.03  & 0.13 \\
\hline
\end{tabular}
\caption{Fitting results for the sample of FSRQs.}
\label{tab:results_fsrq}
\end{table}

\begin{figure}
   \centering
   \includegraphics[width=\hsize]{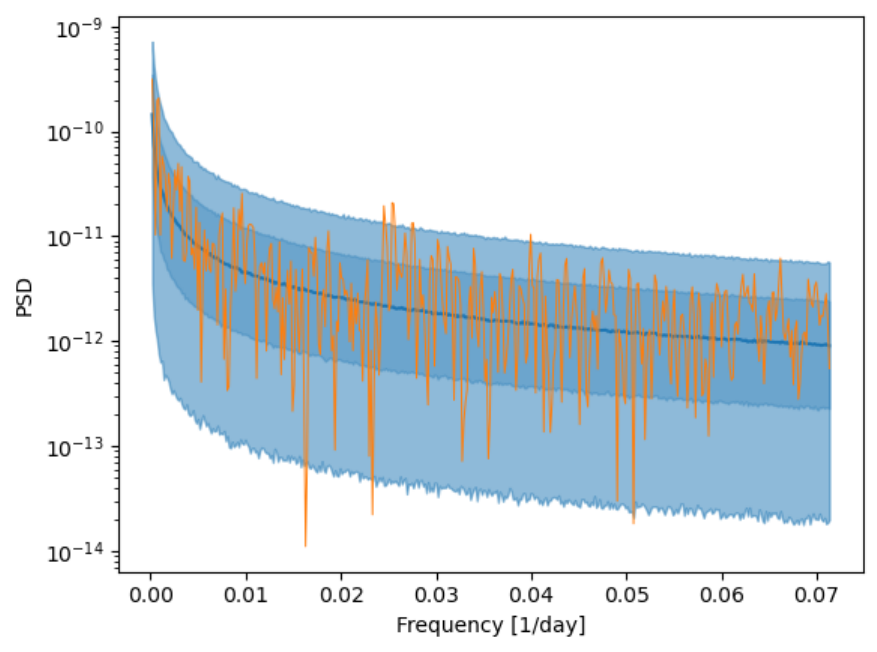}
   \caption{Periodogram of 3C 279 overlaid with the best-fit PSD model and 68\% and 95\% quantile envelopes.}
   \label{fig:psd_3c}
\end{figure}

\begin{figure}
   \centering
   \includegraphics[width=\hsize]{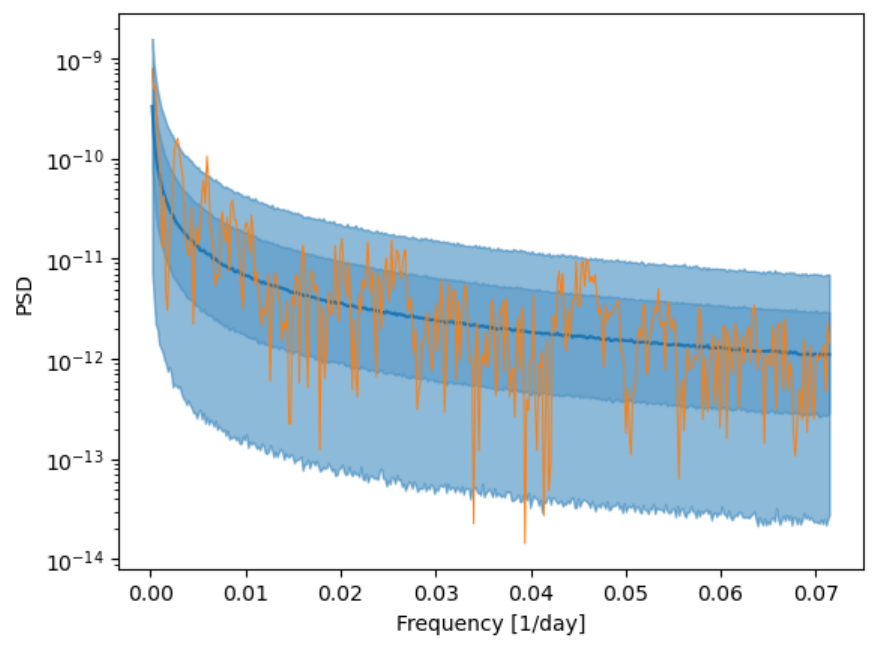}
   \caption{Periodogram of CTA 102 overlaid with the best-fit PSD model and 68\% and 95\% quantile envelopes.}
   \label{fig:psd_cta102}
\end{figure}

\begin{figure}
   \centering
   \includegraphics[width=\hsize]{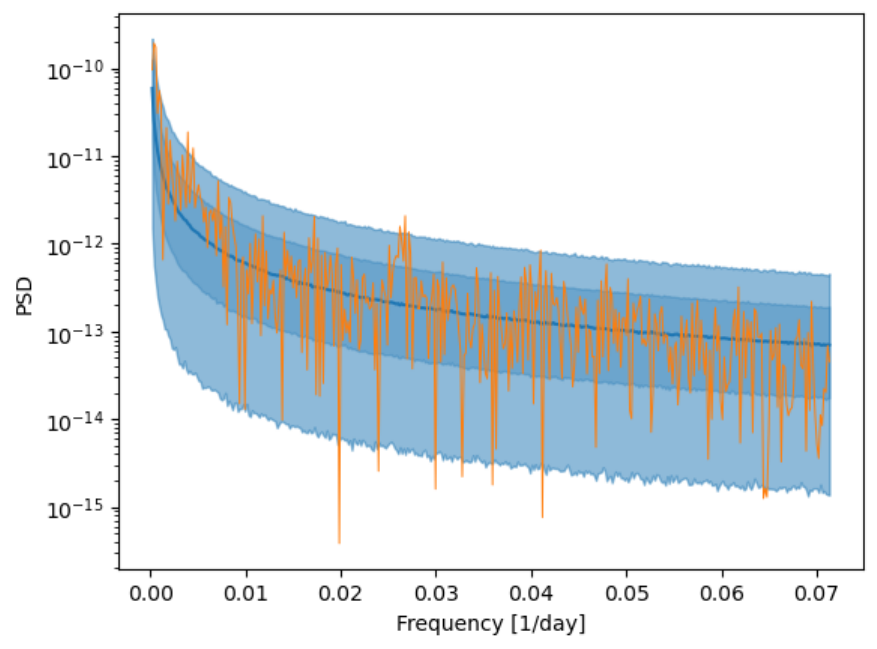}
   \caption{Periodogram of 4C +01.02 overlaid with the best-fit PSD model and 68\% and 95\% quantile envelopes.}
   \label{fig:psd_4c}
\end{figure}

The power spectral density analysis of the three selected flat-spectrum radio quasars reveals a consistent pattern of variability across the sample. The spectral indices obtained for the PSDs of 3C 279, CTA 102, and 4C 01.02 fall within a narrow range between $-0.84$ and $-1.01$, indicating moderate stochastic variability in these sources. A notable feature in the observed PSDs of these FSRQs is the presence of a Poissonian white noise component at high frequencies, similar to what was observed for the Markarian sources. This white noise contribution limits the extent to which a single power-law model can fully describe the underlying variability process. The PSD flattening at high frequencies suggests that a more complete modeling approach, incorporating an additive white noise component alongside the red-noise power law, could improve the fit and provide a more precise characterization of the variability. The fitted model and the statistical envelopes are shown in figure~\ref{fig:psd_3c}, ~\ref{fig:psd_cta102}, ~\ref{fig:psd_4c}.


\begin{figure}
   \centering
   \includegraphics[width=\hsize]{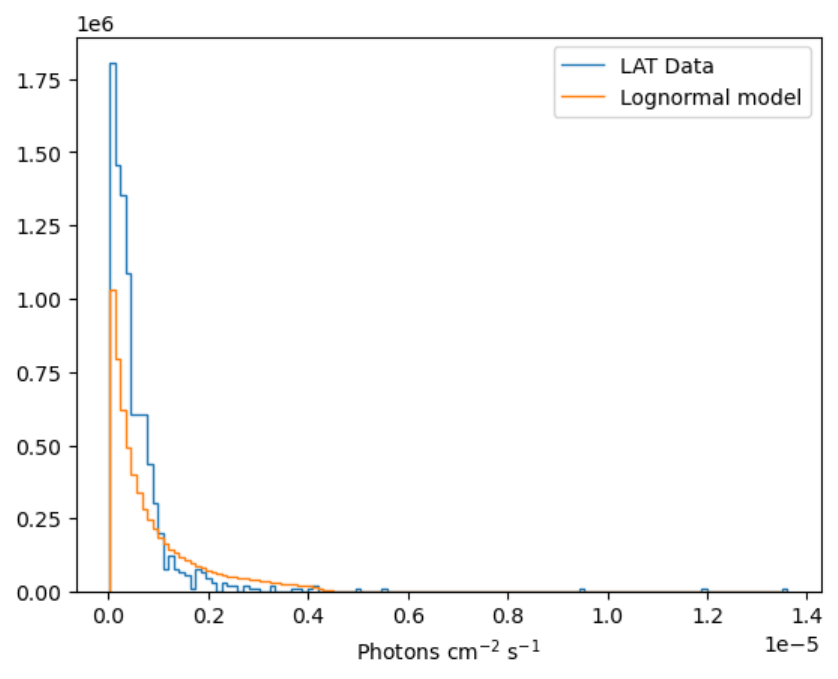}
   \caption{Flux distribution of 3C 279 compared to the best-fit lognormal model.}
   \label{fig:pdf_3c}
\end{figure}

\begin{figure}
   \centering
   \includegraphics[width=\hsize]{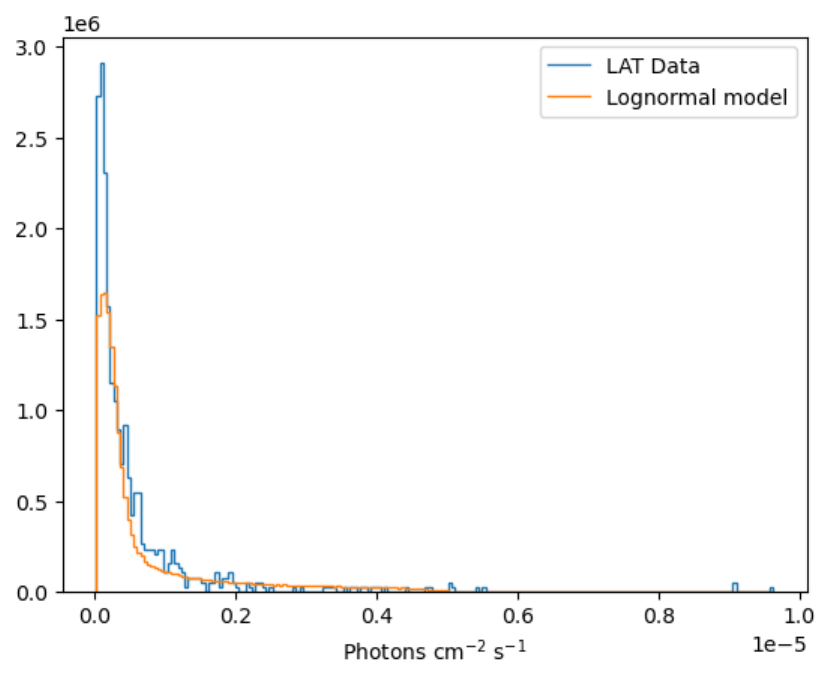}
   \caption{Flux distribution of CTA 102 compared to the best-fit lognormal model.}
   \label{fig:pdf_cta102}
\end{figure}

\begin{figure}
   \centering
   \includegraphics[width=\hsize]{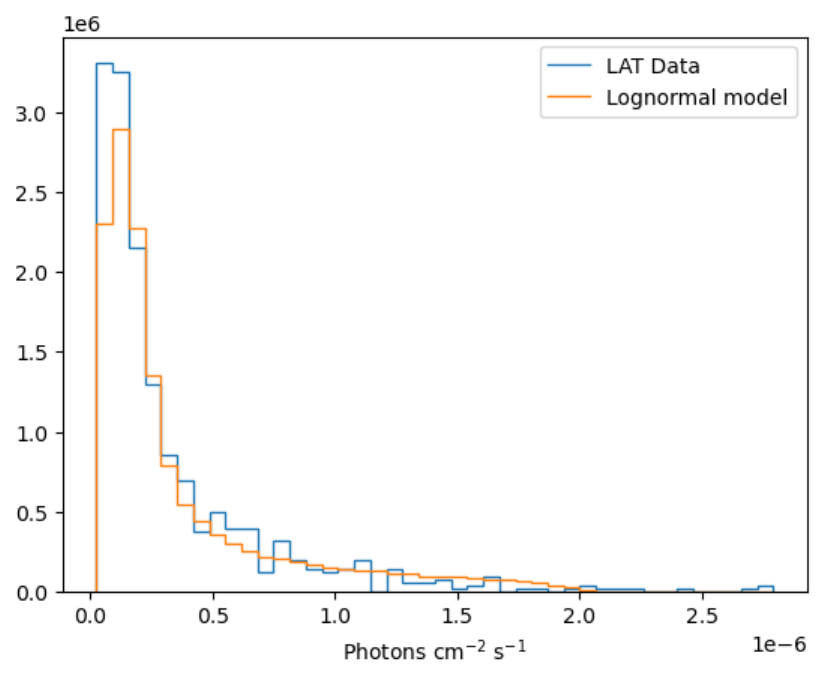}
   \caption{Flux distribution of 4C +01.02 compared to the best-fit lognormal model.}
   \label{fig:pdf_4c}
\end{figure}

On the side of amplitude distributions, all three FSRQs exhibit highly peaked flux distributions that deviate significantly from Gaussianity and also from a lognormal model. The extreme concentration of flux values around the skewed peak, combined with very sparse but extended tails limits the effectiveness of the simple PDF fitting techniques, making it challenging to distinguish between different standard statistical models. This is reflected in the results of the comparison of the lognormal hypothesis against the normal, which yields significances of 1.5 to 1.75 $\sigma$ for all three objects. The fitted PDFs are shown in figure~\ref{fig:pdf_3c}, ~\ref{fig:pdf_cta102}, ~\ref{fig:pdf_4c}.


The complete results for the PSD and PDF fitting for the three FSRQ sources are shown in table~\ref{tab:results_fsrq}

\section{Conclusions}

This work introduces \texttt{gammapy\_SyLC}, a Python package designed to bridge the gap between theoretical models and observational data in time-domain astrophysics. By implementing optimized versions of the Timmer \& Koenig and Emmanoulopoulos algorithms, the package enables the generation of synthetic light curves with specified power spectral densities and amplitude distributions. Additionally, its  PSD and PDF fitting functionalities allow users to extract physical parameters and optimal models from observed light curves, as demonstrated through the analysis of Fermi-LAT gamma-ray light curves.

Looking ahead, there are several opportunities to enhance and expand \texttt{gammapy\_SyLC}. Future developments could include more complex composite models for PSD and PDF fitting, improvements in computational efficiency for high-cadence light curves, and the direct interface with \texttt{gammapy}, offering users the possibility to obtain the results of \texttt{gammapy\_SyLC} algorithms directly as \texttt{gammapy} objects. 

As new data from upcoming observatories become available, such as the Cherenkov Telescope Array Observatory (CTAO), the package could play an important role by giving users simple ways to interface themselves with time-domain modeling and fitting.

\begin{acknowledgements}
The author thanks Catherine Boisson for the great support and guidance throughout all the phases of this work.  The author is also grateful to the gammapy dev team, in particular Bruno Khélifi, Régis Terrier, Kirsty Feijen, Maxime Regeard, and Atreyee Sinha. Thanks also to the CTAO AGN group, especially Jonathan Biteau, Jean-Philippe Lenain, Lea Heckmann, and Andreas Zech for the suggestions, appreciation of this work, and push to improve it. A special thank you to Antonio Condorelli for the fruitful discussions.
\end{acknowledgements}


\bibliographystyle{aa} 
\bibliography{bibliography} 

\begin{thebibliography}{24}
\expandafter\ifx\csname natexlab\endcsname\relax\def\natexlab#1{#1}\fi

\bibitem[{Abdo {et~al.}(2011{\natexlab{a}})}]{Abdo2011_Mrk501}
Abdo, A.~A. {et~al.} 2011{\natexlab{a}}, ApJ, 727, 129

\bibitem[{Abdo {et~al.}(2011{\natexlab{b}})}]{Abdo2011}
Abdo, A.~A. {et~al.} 2011{\natexlab{b}}, ApJ, 736, 131

\bibitem[{Acciari {et~al.}(2020)}]{Acciari2020}
Acciari, V.~A. {et~al.} 2020, MNRAS, 492, 5354

\bibitem[{Ackermann {et~al.}(2011)}]{Ackermann2011}
Ackermann, M. {et~al.} 2011, ApJ, 743, 171

\bibitem[{Ackermann {et~al.}(2015)}]{Ackermann2015}
Ackermann, M. {et~al.} 2015, ApJ, 813, L41

\bibitem[{Ackermann {et~al.}(2016)}]{Ackermann2016}
Ackermann, M. {et~al.} 2016, ApJ, 824, L20

\bibitem[{Ahnen {et~al.}(2017)}]{Ahnen2017}
Ahnen, M.~L. {et~al.} 2017, \aap, 620, A181

\bibitem[{Ahnen {et~al.}(2018)}]{Ahnen2018}
Ahnen, M.~L. {et~al.} 2018, \aap, 620, A181

\bibitem[{Albert {et~al.}(2007)}]{Albert2007}
Albert, J. {et~al.} 2007, ApJ, 669, 862

\bibitem[{Aleksić {et~al.}(2015)}]{Aleksic2015}
Aleksić, J. {et~al.} 2015, \aap, 585, A76

\bibitem[{Biteau \& Giebels(2012)}]{Biteau2012}
Biteau, J. \& Giebels, B. 2012, \aap, 548, A123

\bibitem[{Donath {et~al.}(2023)}]{donath2023}
Donath, A. {et~al.} 2023, A\&A

\bibitem[{Emmanoulopoulos {et~al.}(2013)Emmanoulopoulos, McHardy, \& Papadakis}]{emmanoulopoulos2013}
Emmanoulopoulos, D., McHardy, I.~M., \& Papadakis, I.~E. 2013, MNRAS, 433, 907

\bibitem[{Gasparyan {et~al.}(2018)}]{Gasparyan2018}
Gasparyan, S. {et~al.} 2018, Galaxies, 6, 135

\bibitem[{Hayashida {et~al.}(2015)}]{Hayashida2015}
Hayashida, M. {et~al.} 2015, ApJ, 807, 79

\bibitem[{Madejski {et~al.}(1999)}]{Madejski1999}
Madejski, G.~M. {et~al.} 1999, ApJ, 521, 145

\bibitem[{Massaro {et~al.}(2015)}]{Massaro2015}
Massaro, F. {et~al.} 2015, ApJS, 217, 2

\bibitem[{Mcculloch(1986)}]{mcculloch1986}
Mcculloch, J. 1986, Communications in Statistics - Simulation and Computation - CSSC, 15

\bibitem[{Neyman(1937)}]{neyman1937}
Neyman, J. 1937, Philosophical Transactions of the Royal Society of London. Series A, Mathematical and Physical Sciences, 236, 333

\bibitem[{Nolan {et~al.}(2012)}]{Nolan2012}
Nolan, P.~L. {et~al.} 2012, ApJS, 199, 31

\bibitem[{Raiteri {et~al.}(2013)}]{Raiteri2013}
Raiteri, C.~M. {et~al.} 2013, MNRAS, 436, 1530

\bibitem[{Shukla {et~al.}(2018)}]{Shukla2018}
Shukla, A. {et~al.} 2018, MNRAS, 474, 5032

\bibitem[{Timmer \& K{\"o}nig(1995)}]{timmer1995}
Timmer, J. \& K{\"o}nig, M. 1995, \aap, 300, 707

\bibitem[{Virtanen {et~al.}(2020)}]{virtanen2020}
Virtanen, P. {et~al.} 2020, Nature Methods, 17 [\eprint{arXiv:1907.10121}]

\end{thebibliography}


\end{document}